\documentclass[sigconf]{acmart}

\AtBeginDocument{%
  \providecommand\BibTeX{{%
    \normalfont B\kern-0.5em{\scshape i\kern-0.25em b}\kern-0.8em\TeX}}}

\copyrightyear{2024}
\acmYear{2024}
\setcopyright{acmlicensed}

\acmBooktitle{Proceedings of the ACM Web Conference 2024 (WWW '24), May 13--17, 2024, Singapore, Singapore}
\acmISBN{978-1-4503-XXXX-X/18/06}
\acmDOI{XXXXXXX.XXXXXXX}
\usepackage{multirow}
\usepackage{graphicx}
\usepackage{subcaption}
\usepackage{bm}
\usepackage{balance}
\usepackage{hyperref}

\begin{document}
\newsavebox\CBox
\def\textBF#1{\sbox\CBox{#1}\resizebox{\wd\CBox}{\ht\CBox}{\textbf{#1}}}
\title{Leave No One Behind: Online Self-Supervised Self-Distillation for Sequential Recommendation}


\author{Shaowei Wei}
\affiliation{%
  \institution{Ant Group}
  \city{Hangzhou}
  \country{China}}
\email{weishaowei.wsw@antgroup.com}
\orcid{0000-0002-1275-0015}

\author{Zhengwei Wu}
\affiliation{%
  \institution{Ant Group}
  \city{Hangzhou}
  \country{China}}
\email{zejun.wzw@antgroup.com}
\orcid{0000-0002-9695-3863}

\author{Xin Li}
\affiliation{%
  \institution{Ant Group}
  \city{Hangzhou}
  \country{China}}
\email{lixin324625@antgroup.com}
\orcid{0009-0007-1311-0398}

\author{Qintong Wu}
\affiliation{%
  \institution{Ant Group}
  \city{Hangzhou}
  \country{China}}
\email{qintong.wqt@antgroup.com}
\orcid{0000-0002-5885-3984}

\author{Zhiqiang Zhang}
\affiliation{%
  \institution{Ant Group}
  \city{Hangzhou}
  \country{China}}
\email{lingyao.zzq@antgroup.com}
\orcid{0000-0002-2321-7259}

\author{Jun Zhou}
\affiliation{%
  \institution{Ant Group}
  \city{Hangzhou}
  \country{China}}
\authornote{Corresponding author.}
\email{jun.zhoujun@antgroup.com}
\orcid{0000-0001-6033-6102}

\author{Lihong Gu}
\affiliation{%
  \institution{Ant Group}
  \city{Hangzhou}
  \country{China}}
\email{lihong.glh@antgroup.com}
\orcid{0000-0002-0706-3448}

\author{Jinjie Gu}
\affiliation{%
  \institution{Ant Group}
  \city{Hangzhou}
  \country{China}}
\email{jinjie.gujj@antgroup.com}
\orcid{0000-0001-7596-4945}

\renewcommand{\shortauthors}{Shaowei Wei et al.}

\begin{abstract}
Sequential recommendation methods play a pivotal role in modern recommendation systems.
A key challenge lies in accurately modeling user preferences in the face of data sparsity. 
To tackle this challenge, recent methods leverage contrastive learning (CL) to derive self-supervision signals by maximizing the mutual information of two augmented views of the original user behavior sequence. 
Despite their effectiveness, CL-based methods encounter a limitation in fully exploiting self-supervision signals for users with limited behavior data, as users with extensive behaviors naturally offer more information.  
To address this problem, we introduce a novel learning paradigm, named Online Self-Supervised Self-distillation for Sequential Recommendation ($S^4$Rec), effectively bridging the gap between self-supervised learning and self-distillation methods.
Specifically, we employ online clustering to proficiently group users by their distinct latent intents.
Additionally, an adversarial learning strategy is utilized to ensure that the clustering procedure is not affected by the behavior length factor. 
Subsequently, we employ self-distillation to facilitate the transfer of knowledge from users with extensive behaviors (teachers) to users with limited behaviors (students).
Experiments conducted on four real-world datasets validate the effectiveness of the proposed method.
\end{abstract}





\maketitle

\section{INTRODUCTION}
\label{sec:intro}

\begin{figure}[t!]	
	\centering
	\begin{subfigure}{0.47\columnwidth}
		\centering
		\includegraphics[width=1\columnwidth]{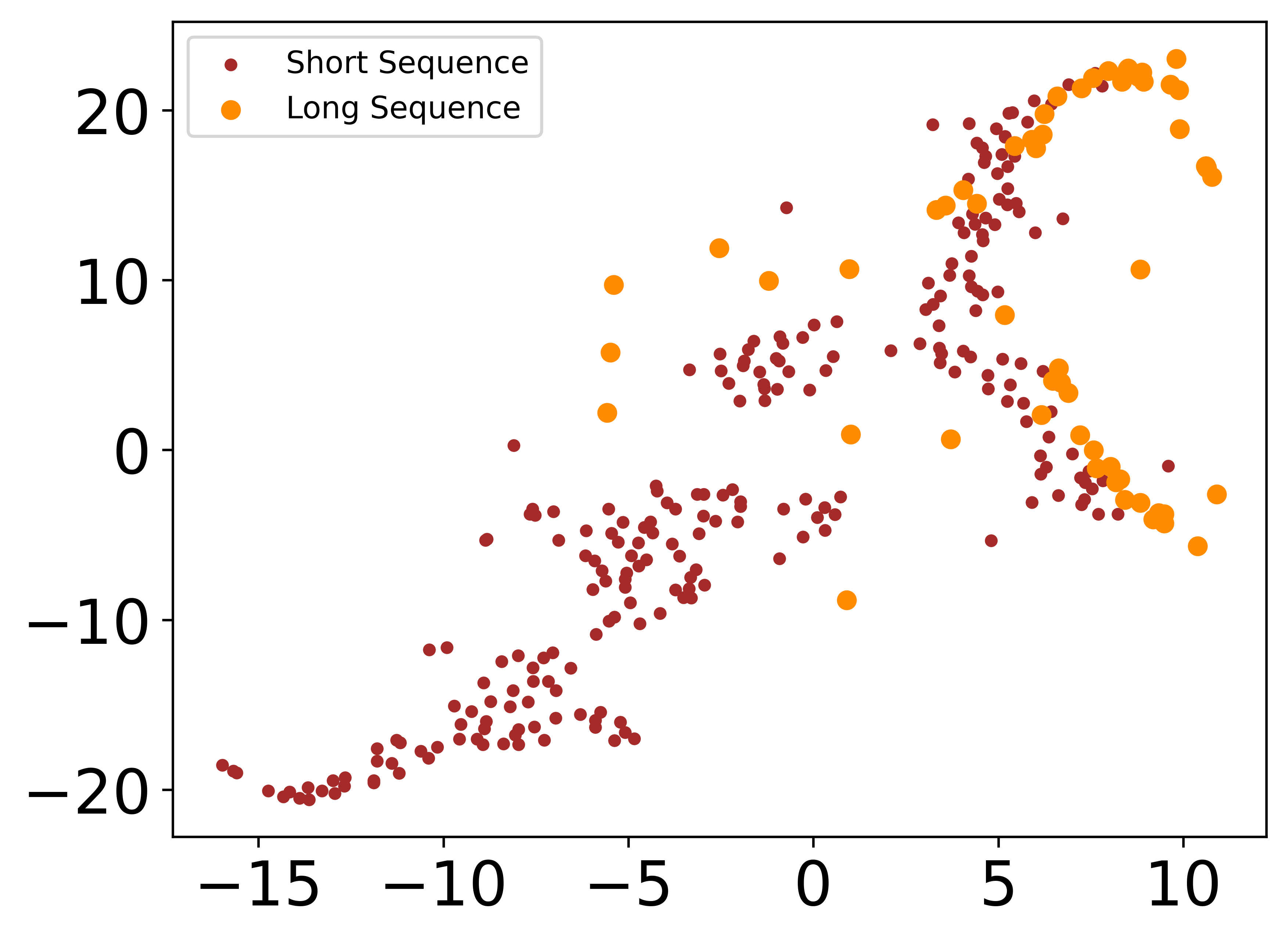}
		\caption{Sequence level}
	\end{subfigure}
	\quad
	\begin{subfigure}{0.47\columnwidth}
		\centering
		\includegraphics[width=1\columnwidth]{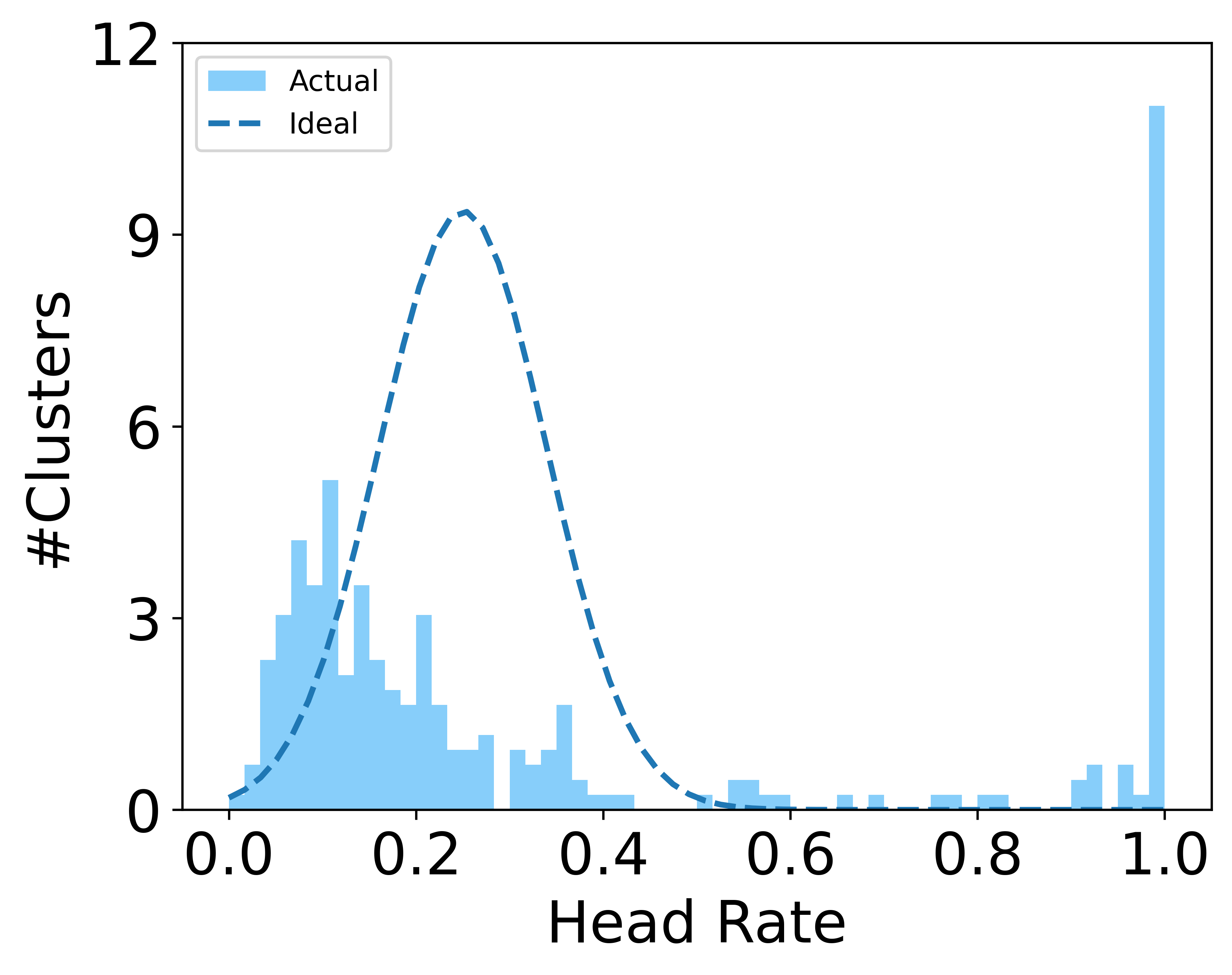}
		\caption{Cluster level}
	\end{subfigure}
	\caption{Visualization of clustering for sequence granularity and cluster granularity on an amazon dataset. The horizontal and vertical axes of Fig.1(a) represent the two-dimensional spatial coordinates of the user sequence embedding vector using the t-SNE dimensionality reduction technique.}
    \label{fig:intro_illustration} 
\end{figure}

As an important recommendation paradigm, sequential recommendation has been playing a vital role in online platforms, e.g., Amazon and Alibaba. 
Generally, sequential recommendation takes a sequence of user-item interactions as the input and aims to predict the subsequent user-item interactions that may happen in the near future through modelling the complex sequential dependencies embedded in the sequence of historical interactions. 
Early works based on Markov Chains~\cite{rendle2010factorizing,he2016fusing} focus on modelling simple low-order sequential dependencies. Afterward, deep learning networks, such as recurrent neural networks (RNN)~\cite{hidasi2015session,hidasi2018recurrent}, convolutional neural networks (CNN)~\cite{tang2018personalized,yuan2019simple}, and memory networks~\cite{huang2018improving} have drawn attention for sequential recommendations due to the powerful non-linear expressive capacity. In addition, transformer-based~\cite{vaswani2017attention,kang2018self,sun2019bert4rec} models have gained popularity for sequential recommendations. They can effectively learn users' preferences by estimating an importance weight for each item.

Although these methods have achieved promising results, they usually only utilize the item prediction task to optimize a huge amounts of parameters, which suffers from data sparsity problem easily. To tackle the problem, inspired by the successes of self-supervised learning in computer vision (CV)~\cite{ChenC23} and natural language processing (NLP)~\cite{GaoYC21}, recent works attempt to use self-supervised learning techniques to optimize the user representation model for improving sequential recommendation systems. These methods typically derive self-supervision signals through maximizing the mutual information of two augmented views of the original user behavior sequence.

Despite their effectiveness, aforementioned methods fail to further extract supervision information across historical interactions. In practice, users consume each item based on their latent intents, which can be perceived as a subjective motive for their interaction. This motivates the exploration~\cite{MaZYCW020,CoSeRec2021} to extract shared underlying intents among users, which can be utilized to guide the recommendation system in providing more relevant recommendations. Since these methods require labels to model the user's intents, ICLRec~\cite{iclrec2022} learns users’ underlying intent distributions from all user interaction sequences via clustering.  However, clustering algorithms typically involve operations over entire datasets, which can be computationally challenging and less efficient dealing with large-scale datasets.

Furthermore, these methods also encounter a limitation in fully exploiting self-supervision signals for users with limited behavior data, as users with extensive behaviors naturally offer
more information. As illustrated in Figure~\ref{fig:intro_illustration}, the learned representations of users with extensive behaviors (long sequences) tend to be clustered by themselves which are relatively separated from users with limited behaviors (short sequences). However, the learned user representations should be affected by users' latent intents and unaffected by the observed sparsity of behavior sequence. Many studies point that uniform representation distribution is a crucial factor for the performance of contrastive learning methods~\cite{wang2020understanding,yu2022graph}. Previous CL-based and intention modelling methods fail to handle the distribution discrepancy between these two types of users, which hinders the sequence recommendation performance, especially for the users with limited behaviors.

To address these problems, we introduce a novel learning paradigm, named Online Self-Supervised Self-distillation for Sequential Recommendation ($S^4$Rec), effectively bridging the gap between self-supervised learning and self-distillation methods.
Specifically, we employ online clustering to proficiently group users by their distinct latent intents. Additionally, an adversarial learning strategy is utilized to ensure that the clustering procedure is not affected by the behavior length factor. 
Subsequently, we employ self-distillation to facilitate the transfer of knowledge from users with extensive behaviors (teachers) to users with limited behaviors (students).

The main contributions of this paper are summarized as follows:
\begin{itemize}
    \item We propose a novel learning paradigm for sequential recommendation, which bridges the gap between self-supervised learning and self-distillation methods. To the best of our knowledge, this is the first work to apply self-distillation techniques to the sequential recommendation.
    \item We propose online clustering and adversarial learning modules to learn user representation clusters which are unaffected by the sparsity of behavior. Based on the learned clusters, the cluster-aware self-distillation module is employed to transfer knowledge from users with extensive behaviors to users with limited behaviors.
    \item Extensive experiments are conducted on four real-world datasets, which show the state-of-the-art performance of the proposed $S^4$Rec model.
\end{itemize}
\section{RELATED WORK}
\subsection{Self-Supervised Learning for Recommendation}
Self-Supervised Learning (SSL) become prevalent in different research areas, including computer vision \cite{ChenC23}, natural language processing \cite{GaoYC21}, and more. The main target of SSL is to capture high-quality and information-rich representations through the feature itself. There have also been some recent works to apply SSL to sequential recommendations. 
For example, S$^{3}$-Rec \cite{ZhouWZZWZWW20}  adopts a pre-training and fine-tuning strategy with four self-supervised tasks, and first proposes to maximize the  mutual information between historical items and their attributes. 
COTREC~\cite{SSGraph21} introduces a graph-based recommendation model that utilizes the session-based graph to augment two views, exhibiting the internal and external connectivities of sessions, thereby supervising each other through contrastive learning.
While DCN~\cite{DualContrastiveSR22} broadens its scope by incorporating user sequence data on the item side and construct a dual contrastive learning network to enhance recommendation performance, our paper narrows its focus to user-side modeling.
CL4SRec \cite{XieSLWGZDC22} introduces three data-level augmentation approaches (crop/mask/reorder, referred to as invasive augmentation methods in the paper) to structure positive views.
Later, CoSeRec \cite{CoSeRec2021} aims to produce robust augmented sequences based on item denpendencies since random item perturbations may weaken the confidence of positive pairs. 
ICLRec \cite{iclrec2022} conducts clustering among all user behavior sequences to obtain user's intent, and optimizes sequential recommendation model by maximizing the  mutual information between sequence and corresponding intentions.

\subsection{Intention Learning for Recommendation}
Many approaches have been proposed to study users' intents behind each user's behavior for improving recommendations \cite{MaZYCW020,chen2020improving,tan2021sparse}.
DSSRec \cite{MaZYCW020} introduces a sequence2sequence training strategy to capture extra supervision information. An intent variable is employed to extract mutual information between an individual user's past and future interaction sequences. 
ICLRec \cite{iclrec2022} learns users' intent distribution from unlabeled user behavior sequences and optimize SR models with contrastive learning  by considering the obtained intents.
ISRec \cite{LiWZMC023} extracts intentions of target user from sequential contexts, then takes complex intent transition into account through the message-passing mechanism on intention graphs. 

\subsection{Sequential Recommendation}
Sequential recommendation aims to learn users' interests and forecast the next items they would most like to interact with  by modeling the sequences of their historical interactions. 
Early works based on Markov Chains \cite{rendle2010factorizing,he2016fusing} focus on modeling simple low-order sequential dependencies. 
Afterward, deep learning networks, such as recurrent neural networks (RNN)\cite{hidasi2015session,hidasi2018recurrent}, convolutional neural networks (CNN)\cite{tang2018personalized,yuan2019simple} and memory networks \cite{huang2018improving} have drawn attention for sequential recommendations due to the powerful nonlinear expressive capacity. 
Recently, transformer-based \cite{vaswani2017attention} models have gained popularity for sequential recommendations. 
Typically, SASRec \cite{kang2018self} uses self-attention mechanism to dynamically assign weights to each item. 
BERT4Rec \cite{sun2019bert4rec} proposes a deep bidirectional transformer model to extract both left and right-side behaviors information. 
ASReP \cite{liu2021augmenting} further solves data sparsity problem by introducing a pretrained transformer on the revised interaction sequences to augment short sequences. 
\section{PRELIMINARIES}
\subsection{Problem Formulation}
We denote $\mathcal{U}$ and $\mathcal{I}$ as the user set and item set, respectively. For each user $u\in \mathcal{U}$, his/her chronological interaction sequence can be represented as $\mathcal{S}_{u}=[s^{1}_{u},...,s^{l}_{u},...,s^{L}_{u}]$, where $s^{l}_{u}$ denotes the 
$l$-th item that user $u$ interacted and $L$ is the maximum sequence length. The goal of sequential recommendation is to predict the next item $s^{L+1}_{u}$ which the user $u$ will most likely interact with given the  behavior sequence $\mathcal{S}_{u}$. To this end, the classical objective function for SR is usually formalized as follows:
\begin{equation}
\small
\centering
\begin{split}
\mathcal{L}_{SR}=\sum_{u=1}^{|\mathcal{U}|}\sum_{l=2}^{L}-\log p_{\theta}(s^{l+1}_{u}|s^{1}_{u},s^{2}_{u},...,s^{l}_{u}),
\end{split}
\label{L_sr}
\end{equation}
where $\theta$ is the parameters of a neural network $f_{\theta}$ that encodes sequential feature into latent vectors: $\mathbf{z}_{u}=f_{\theta}(\mathcal{S}_{u})$. The probability $p(s^{l+1}_{u}|\mathbf{z}_{u}^{l})$ is computed based on the similarity between the encoded sequential patterns $\mathbf{z}_{u}^{l}$ and the representation of the next item $s^{l+1}_{u}$. In serving stage, the items with the highest probability will be recommended to the user $u$.
\subsection{Sequence Augmentation Operators}
Given an original behavior sequence $\mathcal{S}_{u}$, several random sequence-level augmentation strategies can be employed \cite{CoSeRec2021,XieSLWGZDC22}. Following previous works, we specifically apply \textit{Mask}, \textit{Crop}, \textit{Reorder} and \textit{Insert}. The operator definitions are detailedly listed in Appendix~\ref{sa_operators}.

\subsection{Latent Intent Modeling in SR}
Due to subjective reasons, while users face various items in a recommendation system, they may have multiple intentions (e.g., purchasing outdoor equipment, preparing for lectures, just killing time, etc.). The intent variable can be formed as $\mathbf{\mu}\in \mathbb{R}^{K\times d}$. Then the probability of a user interacting with a certain item can be rewritten as $\mathbb{E}_{\mathbf{\mu}}[p(s^{l+1}_{u}|\mathbf{z}_{u}^{l},\mathbf{\mu})]$. As users intents are usually implicit, some work \cite{iclrec2022}  attempts to infer this 
latent intents by unsupervised approach, such as clustering.
\section{METHODOLOGY}
\subsection{Clustering On The Fly}
Previous work learns users’ implicit intents based on user interaction data typically employ clustering methods, such as ICLRec\cite{iclrec2022}. It firstly encodes all the sequences $\{\mathcal{S}_{u}\}_{u=1}^{|\mathcal{U}|}$ by a sequence encoder $f_{\theta}$. Subsequently, ICLRec executes $K$-means clustering over all the sequence representations $\{\mathbf{z}_{u}\}_{u=1}^{|\mathcal{U}|}$ to obtain cluster assignment $\mathbf{P}\in \mathbb{R}^{|\mathcal{U}|\times K}$. 

However, one main issue of these clustering-based methods is that they do not scale well with the dataset as they require a pass over the entire dataset to capture cluster assignments that are used as targets during training. In addition, there is no correspondence between two consecutive cluster assignments. Hence, the final prediction layer learned from an assignment may become irrelevant for the following one and thus needs to be reinitialized from scratch at each epoch, which considerably disrupts the model training.
In this work, we describe an alternative~\cite{caron2020unsupervised} to mapping sequence representations to prototype latent space on the fly in order to scale to large uncurated datasets, and thus retain correspondence.

Firstly, the original interaction sequence is mapped into a vector representation by an encoder as following:
\begin{equation}
\centering
\begin{split}
\mathbf{z}_{u}=f_{\theta}(\mathcal{S}_{u}),
\end{split}
\label{encoder}
\end{equation}
where $f_{\theta}$ is an alternative sequence encoder, which is set as SASRec \cite{kang2018self} in this paper.

\subsection{Cluster-aware Self-distillation}
Once the prototypes $\bm{\mu}$ and corresponding clustering assignments $\mathbf{p}_u$ are obtained, they are employed to construct the supervisory signals for the self-supervision task. More precisely, we propose cluster-aware two-fold self-distillation (CSD) modules: a sequence-level contrastive module and a cluster-level self-distillation module. 
Concretely, the sequence-level contrastive module maximizes mutual information among the positive augmentation pair of the sequence itself while promoting discrimination ability to the negatives. In parallel, the cluster-level self-distillation module aligns each user's behavior sequence to its corresponding intents consistently. The detail is described as follow.

The distillation loss makes use of intent representation which provides additional supervision signals to the  sequence embedding and endows the generalization ability to infer the next item.

\subsection{Head-tail Adversarial Learning}
\label{adv_learning}
Clustering can easily group long and short sequences into separate clusters, indicating that the clusters possess semantic information regarding the sequence sparsity. While the tail sequences are clustered together, the information beyond the length of the sequence within the cluster would be very sparse, thereby impacting the efficacy of  distillation on the tail sequences.

Taking inspiration from the advancement in generative models \cite{gan2014}, we propose incorporating an additional adversarial task of head-tail classification. This new task aims to promote a more uniform distribution of head and tail sequences across different clusters, ultimately boosting the overall recommendation performance.



With respect to the classifier, the classification loss is minimized by finding the category of sequence embeddings. While for the recommendation model, the classification loss is reversed which pushes sequence embeddings of the same category far from each other and not to form clusters. Meanwhile, the main task of minimizing the recommendation loss forces the learned embedding space to retain interest preference semantics.


With the help of adversarial learning, the impact of the head-tail property sequence would be eliminated to some extent.

\subsection{Multi-task Training}
We adopt a multi-task strategy where the main next-item prediction, the cluster assignment, the cluster-aware self-distillation and the adversarial learning task are jointly optimized. 
\section{EXPERIMENTS}
\subsection{Experimental Settings}
\subsubsection{\textbf{Datasets.}}
We conduct experiments on four widely used benchmark datasets with diverse distributions: \textbf{Beauty}, \textbf{Sports} and \textbf{Toys} are three subcategories constructed from Amazon review datasets~\cite{McAuleyTSH15}; \textbf{ML-1M} is a famous movie rating dataset comprising one million ratings. We pre-process these
datasets in the same manner following ~\cite{kang2018self,tang2018personalized,CoSeRec2021} by removing items and users that occur less than five times. For each dataset, we manually select the longest 20\% sequence as head sequences and the rest as tail sequences. Table \ref{statistics_table} shows dataset statistics after pre-processing.

\subsubsection{ \textbf{Evaluation Metrics.}}
Following previous works ~\cite{sun2019bert4rec,CoSeRec2021,iclrec2022}, we adopt two metrics evaluating the performance
of SR models: top-$k$ Hit Ratio@$k$ (HR@$k$) and top-$k$ Normalized Discounted Cumulative Gain (NDCG@$k$) with $k$ chosen from $\{5, 20\}$. For each user's behavior sequence, we reserve the last two items for validation and test, respectively, and use the rest to train SR models. 

\subsubsection{\textbf{Baseline Models.}}
We compare our proposed $S^{4}$Rec  with three categories of methods:
\begin{itemize}
\item \textbf{Non-sequential models.} BPR-MF~\cite{RendleFGS09} leverages Bayesian inference to provide personalized ranking of items for users based on implicit feedback data.
\item \textbf{Standard sequential models.} Caser ~\cite{tang2018personalized} is a CNN-based approach, GRU4Rec ~\cite{hidasi2015session} is an RNN-based method, and SASRec ~\cite{kang2018self} is one of the state-of-the-art Transformer-based baselines for SR. They optimize the same objective but differ in sequence encoder structures.
\item \textbf{Sequential models considering SSL.} BERT4Rec ~\cite{sun2019bert4rec} proposes a deep bidirectional transformer model to extract both left and right-side behaviors information. S$^{3}$-Rec ~\cite{ZhouWZZWZWW20}  adopts a pre-training and fine-tuning strategy with four self-supervised tasks. CL4SRec ~\cite{XieSLWGZDC22} introduces three data-level augmentation approaches to construct positive views. This line of works all utilize the transformer as sequence encoder but adopt distinct constrastive learning tasks. 
\item \textbf{Sequential models with additional latent factors.} DSSRec ~\cite{MaZYCW020} introduce an intent variable to extract mutual information between an individual user’s past and future interaction sequences. ICLRec ~\cite{iclrec2022} leverages the clustered latent intent factor and contrastive self-supervised learning to optimize SR. 
\end{itemize}
\noindent Implementation details are detailed in Appendix~\ref{implementation_details}.


\begin{table}[t]
\caption{Statistics of used datasets.}
\centering
\begin{tabular}{c c c c c c}
    \toprule
    Datasets &Beauty &Sports &Toys &ML-1M\\
    \hline
    \#Users &22363 &35598 &19412 &6041 \\
    \#Items &12101 &18357 &11924 &3417 \\
    \#Actions &0.2m &0.3m &0.17m &0.99m \\
    Avg.length  &8.9 &8.3 &8.6 &165.5\\
    Sparsity    &99.95\% &99.95\% &99.93\% &95.15\%\\
    \bottomrule
\end{tabular}
\label{statistics_table}
\end{table}

\subsection{Overall Performance}
In Table~\ref{over_all_compare}, we present the consistent performance gain of the proposed $S^4$Rec against baselines on different datasets. 
The major results are summarized as follows:
\begin{table*}[t!]
    \caption{Overall performance. Bold scores represent the highest results of all methods. Underlined scores stand for the second highest. "$*$" denotes the statistical siginificance for $p< 0.01$ compared to the best baseline methods with paired $t$-test.}
    \centering
    \resizebox{2.0\columnwidth}{!}{
    \begin{tabular}{l l|c c c c c c c c c|c}
        \hline
        Dataset&Metric &BPR &GRU4Rec &Caser &SASRec &DSSRec &BERT4Rec &$S^{3}$-Rec &CL4SRec &ICLRec &Ours\\
        \hline
        \multirow{4}[2]{*}{Beauty}
        &HR@5
        &0.0212 &0.0111 &0.0251 &0.0374 &0.0410 &0.0360 &0.0189 &0.0423 &\underline{0.0475} &\textbf{0.0519}* \\
        &HR@20
        &0.0589 &0.0478 &0.0643 &0.0901 &0.0914 &0.0984 &0.0487 &0.0994 &\underline{0.1050} &\textbf{0.1071}* \\
        &NDCG@5
        &0.0130 &0.0058 &0.0145 &0.0241 &0.0261 &0.0216 &0.0115 &0.0281 &\underline{0.0316} &\textbf{0.0348}* \\
        &NDCG@20
        &0.0236 &0.0104 &0.0298 &0.0387 &0.0403 &0.0391 &0.0198 &0.0441 &\underline{0.0478} &\textbf{0.0505}* \\
        \hline
        \multirow{4}[2]{*}{Sports}
        &HR@5
        &0.0141 &0.0162 &0.0154 &0.0206 &0.0214 &0.0217 &0.0121 &0.0217 &\underline{0.0267}  &\textbf{0.0284}* \\
        &HR@20
        &0.0323 &0.0421 &0.0399 &0.0497 &0.0495 &0.0604 &0.0344 &0.0540 &\underline{0.0644} &\textbf{0.0656}* \\
        &NDCG@5
        &0.0091 &0.0103 &0.0114 &0.0135 &0.0142 &0.0143 &0.0084 &0.0137 &\underline{0.0177} &\textbf{0.0181}* \\
        &NDCG@20
        &0.0142 &0.0186 &0.0178 &0.0216 &0.0220 &0.0251 &0.0146 &0.0227 &\underline{0.0283} &\textbf{0.0292}* \\
        \hline
        \multirow{4}[2]{*}{Toys}
        &HR@5
        &0.0120 &0.0097 &0.0166 &0.0463 &0.0502 &0.0274 &0.0143 &0.0526 &\underline{0.0571} &\textbf{0.0586}* \\
        &HR@20
        &0.0312 &0.0301 &0.0420 &0.0941 &0.0975 &0.0688 &0.0235 &0.1038 &\underline{0.1110} &\textbf{0.1148}* \\
        &NDCG@5
        &0.0082 &0.0059 &0.0107 &0.0306 &0.0337 &0.0174 &0.0123 &0.0362 &\underline{0.0392} &\textbf{0.0407}* \\
        &NDCG@20
        &0.0136 &0.0116 &0.0179 &0.0441 &0.0471 &0.0291 &0.0162 &0.0506 &\underline{0.0545} &\textbf{0.0565}* \\
        \hline
        \multirow{4}[2]{*}{ML-1M}
        &HR@5
        & 0.0467 & 0.1412 & 0.1331 & 0.1444 & 0.1219 & 0.1142 & \textbf{0.1579} & 0.1520 & 0.1482 & \underline{0.1557}* \\
        &HR@20
        & 0.1295 & 0.3379 & 0.3187 & 0.3337 & 0.2855 & 0.2921 & 0.3435 & \underline{0.3537} & 0.3431 & \textbf{0.3547}* \\
        &NDCG@5
        & 0.0295 & 0.0916 & 0.0845 & 0.0939 & 0.0798 & 0.0713 & \underline{0.0982} & 0.0969 & 0.0964 & \textbf{0.1012}* \\
        &NDCG@20
        & 0.0524 & 0.1469 & 0.1370 & 0.1476 & 0.1257 & 0.1213 & \underline{0.1510} & 0.1539 & 0.1513 & \textbf{0.1570}* \\
        \hline
\end{tabular}}
\label{over_all_compare}
\end{table*}

\subsection{Ablation Study}
Our proposed method contains two essential modules: cluster-aware two-fold self-distillation (CSD) module and Gradient Reversal (GR) layer as an adversarial learning module. To understand the impact of the sub-modules of $S^4$Rec, we conduct ablation study by removing different sub-module. The results reported in Table~\ref{ablation} are based on experiments conducted in the Amazon Beauty dataset. Similar results are also achieved in other datasets.

\begin{table}[t]
\caption{Ablation study of $S^4$Rec on the Beauty dataset.}
\centering  
\begin{tabular}{@{}l|llll@{}}
\toprule
& HR@5          & HR@20           & NDCG@5          & NDCG@20         \\ \hline
SR only         & 0.0374          & 0.0901          & 0.0241          & 0.0387          \\ 
SR+CSD          & 0.0499          & 0.1041          & 0.0332          & 0.0485          \\ 
SR+CSD+GR       & 0.0517          & 0.1044          & 0.0349          & 0.0497          \\ \hline
$S^4$Rec w/o CSD    & 0.0382          & 0.0922          & 0.0258          & 0.0397          \\ 
$S^4$Rec w/o GR     & 0.0487          & 0.1035          & 0.0329          & 0.0480          \\ \hline
\end{tabular}
\label{ablation}
\end{table}

\subsection{Complexity Analysis}

For $S^4$Rec, the time complexity of its main task, clustering task, self distillation task, and adversarial task are $\mathcal{O}(aL^{2}|\mathcal{U}|d)$, $\mathcal{O}(aK|\mathcal{U}|d)$, $\mathcal{O}(aB|\mathcal{U}|d^2)$ and $\mathcal{O}(aC|\mathcal{U}|d)$, respectively,  where \textit{a} is the number of epoch. Since these tasks are jointly trained in one model, the overall complexity is dominated by the term $\mathcal{O}(aL^{2}|\mathcal{U}|d+aB|\mathcal{U}|d^2)$. On the contrary, ICLRec suffers an extra $\mathcal{O}(ab|\mathcal{U}|Kd)$ time complexity in clustering E-step, where \textit{b} is the maximum iteration number. Training time difference is reported in Appendix~\ref{training_time}.




\subsection{Online A/B Testing}
To further strengthen our findings, we present the results of an online A/B testing conducted on a real-world recommendation system, affecting more than 23 million users from November 2, 2023, to November 6, 2023. On the contrary, due to the entire data iteration, traditional clustering will also occupy a large amount of memory footprints, which is difficult to be deployed for real recommendation scenarios with millions of users.
Comprehensive evaluation is listed in Appendix~\ref{abtest}.
\section{CONCLUSION}
In this paper, we present $S^4$Rec, a practical attempt to address the sparsity problem of behavior for the sequential recommendation, which bridges the gap between self-supervised learning and self-distillation methods.
$S^4$Rec utilizes online clustering and adversarial learning modules to optimize user intention clusters unaffected by the sparsity of behavior. Subsequently, the sequence-level contrastive learning considers negative intents augments the representation express capability, while the cluster-aware self-distillation module transfers knowledge from users with extensive behaviors to users with limited behaviors. Experimental results on benchmark datasets confirm and validate the effectiveness of the proposal. 


\clearpage


\bibliographystyle{ACM-Reference-Format}
\balance
\bibliography{reference.bib}

\appendix
\newpage
\section{Sequence Augmentation Operators}
\label{sa_operators}
Sequence-level augmentations aim to create multiple related views from the original user behaviors. $S^4$-Rec adopts the following choices of augmentation operators:

\noindent \textbf{Mask.} It randomly masks a proportion of items in an original sequence. This mask operation can be formulated as:
\begin{equation}
\centering
\begin{split}
\mathcal{S}_{u}^{Mask}=[\hat{s}^{1}_{u},...,\hat{s}^{l}_{u},...,\hat{s}^{L}_{u}],
\end{split}
\label{Mask}
\end{equation}
where $\hat{s}^{l}_{u}$ represents the masked item if $s^{l}_{u}$ is selected, otherwise $\hat{s}^{l}_{u}=s^{l}_{u}$.

\noindent \textbf{Crop.} It randomly removes a continuous sub-sequence
from positions $l$ to $l+l_{c}$ in $\mathcal{S}_{u}$. The length to crop is set by $l_{c}=\delta*|\mathcal{S}_{u}|$ where empirically $\delta=0.8$. The formulation of the cropped sequence is shown below:
\begin{equation}
\centering
\begin{split}
\mathcal{S}_{u}^{Crop}=[s^{1}_{u},...,s^{l}_{u},s^{l+l_{c}}_{u},....,s^{L}_{u}],
\end{split}
\label{Crop}
\end{equation}

\noindent \textbf{Reorder.} It randomly shuffles a continuous sub-sequence
from positions $l$ to $l+l_{c}$ in $\mathcal{S}_{u}$.  The length to reorder is set by $l_{c}=\delta*|\mathcal{S}_{u}|$ where empirically $\delta=0.2$. The formulation of the reordered sequence is as:
\begin{equation}
\centering
\begin{split}
\mathcal{S}_{u}^{Reorder}=[s^{1}_{u},...,\hat{s}^{l}_{u},...,\hat{s}^{l+l_{c}}_{u},....,s^{L}_{u}],
\end{split}
\label{Reorder}
\end{equation}

\noindent \textbf{Insert.} 
It inserts an item chosen randomly from the interaction histories of other users into a randomly selected position within $\mathcal{S}_{u}$. This operation is employed  repeatedly on the sequence to obtain an augmented view. The augmented sequence could be formulated by:
\begin{equation}
\centering
\begin{split}
\mathcal{S}_{u}^{Insert}=[s^{1}_{u},...,\hat{s}^{1}_{u},...,\hat{s}^{i}_{u},....,s^{L}_{u}].
\end{split}
\label{Insert}
\end{equation}

\section{Implementation Details}
\label{implementation_details}
\hypersetup{
    colorlinks=true,
    linkcolor=blue,
    filecolor=magenta,      
    urlcolor=blue,
    pdftitle={Overleaf Example},
    pdfpagemode=FullScreen,
}
For BPR-MF and GRU4Rec, we use the source code provided by \href{https://github.com/THUwangcy/ReChorus}{Wang et al.} in PyTorch. For Caser, SASRec, BERT4Rec and $S^{3}$Rec, the source code is provided by \href{https://github.com/RUCAIBox/RecBole}{Zhao et al.} in PyTorch. For \href{https://github.com/abinashsinha330/DSSRec}{DSSRec} \href{https://github.com/salesforce/ICLRec}{ICLRec} and CL4SRec, we use the source code provided by their authors. Our method is implemented in PyTorch as well.  For all models, the dimension of embedding is set as 64, and the maximum sequence length is set as 50 for alignment, following previous works ~\cite{kang2018self, sun2019bert4rec, ZhouWZZWZWW20}. For each baseline model, all other hyper-parameters are set following the suggestions from the original papers.


For our proposed $S^{4}$Rec, the optimizer is Adam~\cite{kingma2014adam}, learning rate is $0.001$, batch size $B$ is $512$, dropout rate is $0.5$, number of clusters $K$ is $128$, number of hidden layers is set from $\{1,2,3\}$. Multi-task objective weights $\alpha,\beta_{1},\beta_{2},\lambda \in \{0,0.01,0.1,1.0\}$. The temperature parameters $\tau_1,\tau_2,\tau_3$ are chosen from $\{0.1,1.0\}$.

\section{Training Efficiency on Clustering Algorithm}
\label{training_time}
We have present the performance of replacing online clustering of S4Rec with K-means, i.e. SR+CSD+GR. Here we further report the training time result in Table \ref{train_time}.
\begin{table}[H]
\caption{Training Time (minutes) on GPU Tesla P100.}
\centering
\begin{tabular}{c c c c c c}
    \toprule
     &Beauty &Sports &Toys &ML-1M\\
    \hline
    $S^4$Rec with K-means &62.6 &109.3 &72.6 &53.8 \\
    $S^4$Rec &31.4 &68.2 &48.1 &28.4 \\
    \bottomrule
\end{tabular}
\label{train_time}
\end{table}
We conclude that K-means clustering algorithm suffers nonnegligible training time. On the contrary, online clustering algorithm provides better training efficiency with lossless performance.

\section{A/B Testing Evaluation}
\label{abtest}
We present the results of an online A/B test conducted on a real-world recommendation system. Due to the excessive computational cost of K-means to over 10 million samples, we deploy another well-performing CL4SRec as the A/B test baseline, instead of ICLRec.

As demonstrated in the following table, our method consistently achieves stable and statistically significant improvements in the real-world large-scale recommendation scenario, our method can achieve statistically significant improvements stably.

\begin{table}[H]
    \caption{Online A/B Testing results and offline AUC uplifts.}
    \centering
    \begin{tabular}{l|rrrr}
        \toprule
        {} & {\begin{tabular}[c]{@{}r@{}}Online \\ CTR \\ Improv.\end{tabular}} & {\begin{tabular}[c]{@{}r@{}}Online \\ CVR \\ Improv.\end{tabular}} & {\begin{tabular}[c]{@{}r@{}}Offline \\ CTR AUC \\ Improv.\end{tabular}} & {\begin{tabular}[c]{@{}r@{}}offline \\ CVR AUC \\ Improv.\end{tabular}} \\ 
        \hline
        { Day5} & { 0.09\%} & { 0.42\%} & { 1.38\%} & { 1.49\%} \\ 
        { Day4} & { 1.19\%} & { 1.68\%} & { 1.24\%} & { 2.95\%} \\ 
        { Day3} & { 3.86\%} & { 6.95\%} & { 2.39\%} & { 0.81\%} \\ 
        { Day2} & { 1.19\%} & { 2.43\%} & { 1.57\%} & { 1.29\%} \\ 
        { Day1} & { 1.67\%} & { 5.09\%} & { 3.90\%} & { 1.48\%} \\
        \bottomrule
    \end{tabular}
    \label{abtesting}
\end{table}

\end{document}